\documentclass[prb,letterpaper,superscriptaddress,aps,floatfix,twocolumn]{revtex4-1}
\usepackage{graphicx}
\usepackage{amsmath}
\usepackage{amsfonts}
\usepackage[small,bf]{subfigure}
\newcommand{\beq}{\begin{equation}}
\newcommand{\eeq}{\end{equation}}

\newcommand{\bk}{{{\bf{k}}}}

\newcommand{\bR}{{{\bf{R}}}}

\newcommand{\bB}{{\bf{B}}}

\newcommand{\bb}{{\bf{b}}}

\newcommand{\bj}{{\bf j}}

\newcommand{\beqa}{\begin{eqnarray}}
\newcommand{\eeqa}{\end{eqnarray}}

\newcommand{\bnabla}{{\boldsymbol \nabla}} 
\newcommand{\bsigma}{{\boldsymbol \sigma}}

\newcommand{\bGamma}{{\boldsymbol \Gamma}}

\begin{document}
\title{Topological properties of Dirac and Weyl semimetals}
\author{A.A. Burkov}
\affiliation{Department of Physics and Astronomy, University of Waterloo, Waterloo, Ontario 
N2L 3G1, Canada} 
\affiliation{Perimeter Institute for Theoretical Physics, Waterloo, Ontario N2L 2Y5, Canada}
\date{\today}
\begin{abstract}
This chapter describes topological (Dirac and Weyl) semimetals from the viewpoint of their observable electromagnetic response. 
We argue that this response may be represented by topological terms with unquantized (non-integer) coefficients and make a connection 
with the Luttinger's theorem, which relates the size of the Fermi surface of an ordinary metal to the number of electrons per unit cell. 
We discuss observable transport phenomena, associated with this topological response. 
\end{abstract}
\maketitle
\section*{Key points}
\begin{itemize}
\item Topological semimetals are intermediate phases between insulators with different electronic structure topology.
\item Topological response of Dirac and Weyl topological semimetals may be viewed as a consequence of the chiral anomaly. 
\item Observable manifestations are intrinsic anomalous Hall effect with a noninteger Hall conductance in units of $e^2/h$ per atomic 
plane, negative longitudinal magnetoresistance, scale-dependent quasiballistic transport and extra low-frequency peak in the optical conductivity. 
\end{itemize}
\section*{Introduction}
\label{sec:1}
Solid crystalline materials are typically categorized as either metals or insulators. Metals have a finite conductivity in the limit of zero temperature, while in insulators the conductivity tends to zero in this limit. 
The existence of metals and insulators is a fundamental consequence of quantum mechanics and is one of the basic macroscopic quantum phenomena. 
Whether a given material is a metal or an insulator is determined by a single parameter: the number of valence electrons per unit cell per spin, or 
filling, denoted as $\nu$ henceforth.  
If $\nu$ is not an integer, the material is a metal (unless electron-electron interactions are so strong that a Mott insulator state is formed). 
The main characteristic of a conventional metal is the existence of a Fermi surface: a surface in the crystal momentum space, separating filled and empty states, whose volume is directly proportional to the fractional part of the filling, a statement known as the Luttinger's theorem. 
On the other hand, if $\nu$ is an integer, the material is either an insulator or a compensated semimetal, which has a number of electron and hole
pockets, whose total volume in momentum space, weighted by the sign of the charge carrier (i.e. negative for electrons and positive for holes), sums up 
to zero. Since the existence of a compensated semimetal is not required by the Luttinger's theorem, it may be viewed as accidental, in the sense that 
it may be deformed to an insulator without changing $\nu$. We may then say that a noninteger $\nu$ always corresponds to a metal, while integer $\nu$ always means an insulator. 

This simple picture, which has been around for many decades, was significantly enriched and modified by the recent developments in understanding the role of topology in condensed matter physics. 
In particular, it has been understood that a (semi)metallic state may arise for topological reasons in situations when the Luttinger parameter is an integer, which would normally correspond to an insulator. 
The simplest example of this are the metallic surface states of topological insulators, which arise inevitably as a consequence of the bulk insulator electronic 
structure topology.
Topological semimetals may also exists in stand-alone bulk systems, most naturally in three spatial dimensions (3D). 
Just as ordinary metals may be viewed as intermediate phases between two band insulators, corresponding to different integer values of $\nu$, 
topological semimetals arise as intermediate phases between insulators with different electronic structure topology at a fixed integer value of $\nu$. 
For example, a magnetic Weyl semimetal is an intermediate phase between an ordinary 3D insulator and an integer quantum Hall insulator with the Hall 
conductance of $e^2/h$ per atomic plane. A time-reversal (TR) invariant Weyl semimetal is an intermediate phase between an ordinary insulator and 
a 3D TR-invariant topological insulator (TI), which has an odd number of two-dimensional (2D) Dirac cone surface states. 
A (type-I) Dirac semimetal is an intermediate phase between an ordinary insulator and a weak topological crystalline insulator, protected by a combination 
of TR, inversion, and crystal rotation symmetries. 
A nodal line semimetal is an intermediate phase between an ordinary insulator and a topological crystalline insulator, protected by mirror symmetry. 

Just as an ordinary metal may be characterized by a fractional value of the filling parameter $\nu$, which determines the size of its Fermi surface, topological semimetals may be characterized by topological invariants, which, although not integer, not quantized, and continuously tunable, 
are otherwise robust to interactions and disorder. Integer values of these invariants correspond to insulators, while fractional values correspond to topological 
semimetal phases, intermediate between the insulators. 
In the simplest case of the magnetic Weyl semimetal, this tunable invariant is the Hall conductance per atomic plane in units of $e^2/h$. 
Any noninteger value of this invariant requires Weyl band-touching nodes to be present when the Luttinger parameter $\nu$ is an integer (otherwise the Hall conductance can take any value due to the contribution of the Fermi surface). 
In other types of topological semimetals, the invariants are more subtle, but the idea remains the same: a noninteger value of the invariant requires 
gapless band-touching points or lines to be present. 

\section*{Magnetic Weyl semimetal}
\label{sec:2}
This section explains the connection between unquantized topological invariants and the existence of gapless band-touching points in the Brillouin zone (BZ) 
on the simplest example of a topological semimetal: a magnetic Weyl semimetal with only a pair (the minimal number) of band-touching nodes. 
Such a Weyl semimetal arises as an intermediate phase between a quantum anomalous Hall and an ordinary insulator in 3D. 

A 3D quantum anomalous Hall insulator may be obtained by making a stack of 2D quantum Hall insulators. 
2D quantum anomalous Hall effect (QAHE) arises naturally in a very thin film of a 3D TI material, doped with magnetic 
impurities.
A 3D TR-invariant TI is a bulk insulator with nontrivial electronic structure topology, which leads to gapless metallic surface states, described by a 2D massless Dirac Hamiltonian of the general form 
\beq
\label{eq:1} 
H = \hbar v_F (\hat z \times \bsigma) \cdot \bk, 
\eeq
where $\bsigma$ is the spin degree of freedom, $v_F$ is the Fermi velocity, and $\hat z$ is the normal to the surface. 
The gaplessness of the spectrum of Eq.~\eqref{eq:1} is protected by TR symmetry, which prohibits the ``mass" 
term in the 2D Dirac Hamiltonian. If time reversal symmetry is violated, which may be accomplished by doping the surface with magnetic impurities, the ``mass" term will be allowed and a gap in the 2D Dirac surface state dispersion will be opened. 
\begin{figure}[t]
\vspace{-2cm}
\includegraphics[width=\columnwidth]{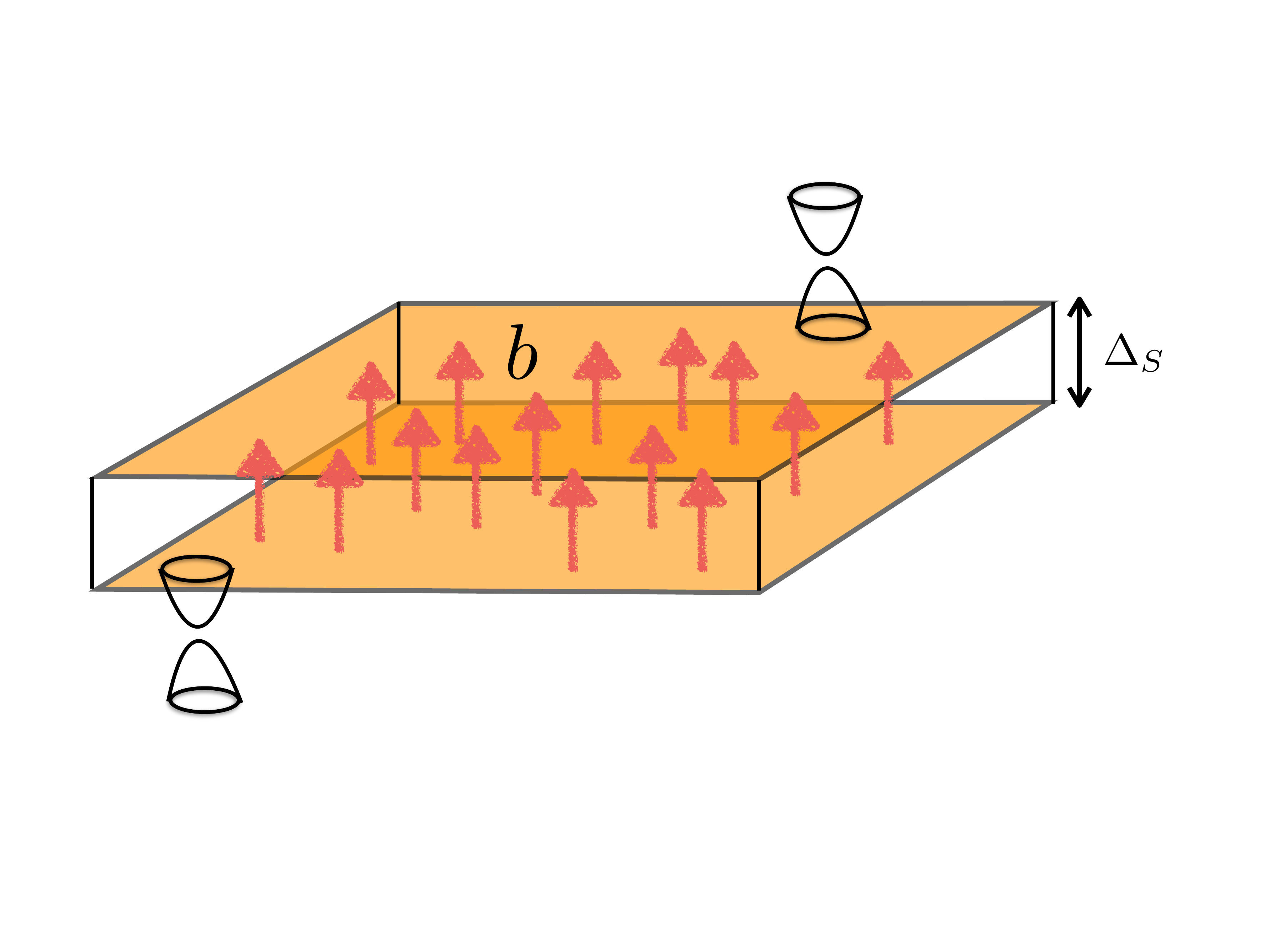}
\vspace{-2cm} 
\caption{Thin film of magnetically doped 3D TI. $\Delta_S$ is the tunneling amplitude between the top and bottom 2D Dirac 
surface states and $b$ is the spin splitting due to magnetized impurities. When $b > \Delta_S$, the film is a quantum anomalous 
Hall insulator with $\sigma_{xy} = e^2/h$. When $b < \Delta_S$, it is an ordinary insulator with zero Hall conductivity.}
\label{fig:1}
\end{figure}

A thin film of magnetically-doped 3D TI, shown in {\bf Figure~\ref{fig:1}},  may be modeled by focusing only on the low-energy degrees of freedom, which 
are simply the 2D Dirac surface states of Eq.~\eqref{eq:1} on the top and bottom surfaces of the film. 
The corresponding Hamiltonian reads
\beq
\label{eq:2}
H = \hbar v_F \tau^z (\hat z \times \bsigma) \cdot \bk + \Delta_S \tau^x + b \sigma^z. 
\eeq
Here the eigenvalues of $\tau^z$ refer to the top or bottom surface degree of freedom, $\Delta_S$ is the probability amplitude for tunneling between the 
top and bottom surfaces of the film, and $b$ is the exchange spin-splitting, which arises due to the presence of 
magnetized impurities. Note that $b \sigma^z$ acts as a ``mass term" for the individual 2D Dirac surface states, as mentioned above. 
A unitary transformation brings this to the form
\beq
\label{eq:3}
H = \hbar v_F (\hat z \times \bsigma) \cdot \bk + (\Delta_S \tau^x + b) \sigma^z, 
\eeq
which may be further brought to a block-diagonal form by diagonalizing the $\Delta_S \tau^x$ matrix
\beq
\label{eq:4}
H_r = \hbar v_F (\hat z \times \bsigma) \cdot \bk + (b + r \Delta_S) \sigma^z, 
\eeq
where $r = \pm$. 
Each of the $2 \times 2$ blocks of Eq.~\eqref{eq:4} is the Hamiltonian of a 2D Dirac fermion with a ``mass" 
$m_r = b + r \Delta_S$. 
The 2D Dirac fermion possesses an important property, which is known as the ``parity anomaly".
In our context this means that $H_{\pm}$ is associated with a Hall conductivity
\beq
\label{eq:5}
\sigma^r_{xy} = \frac{e^2}{2 h} \textrm{sign}(m_r),
\eeq
when the Fermi energy is in the gap between the positive and negative energy bands, obtained by diagonalizing $H_r$
\beq
\label{eq:6}
\epsilon_{r s} (\bk) = s \sqrt{\hbar ^2 v_F^2 \bk^2 + m_r^2}, 
\eeq 
with $s = \pm$.  
\begin{figure}[t]
\vspace{-2cm}
\includegraphics[width=\columnwidth]{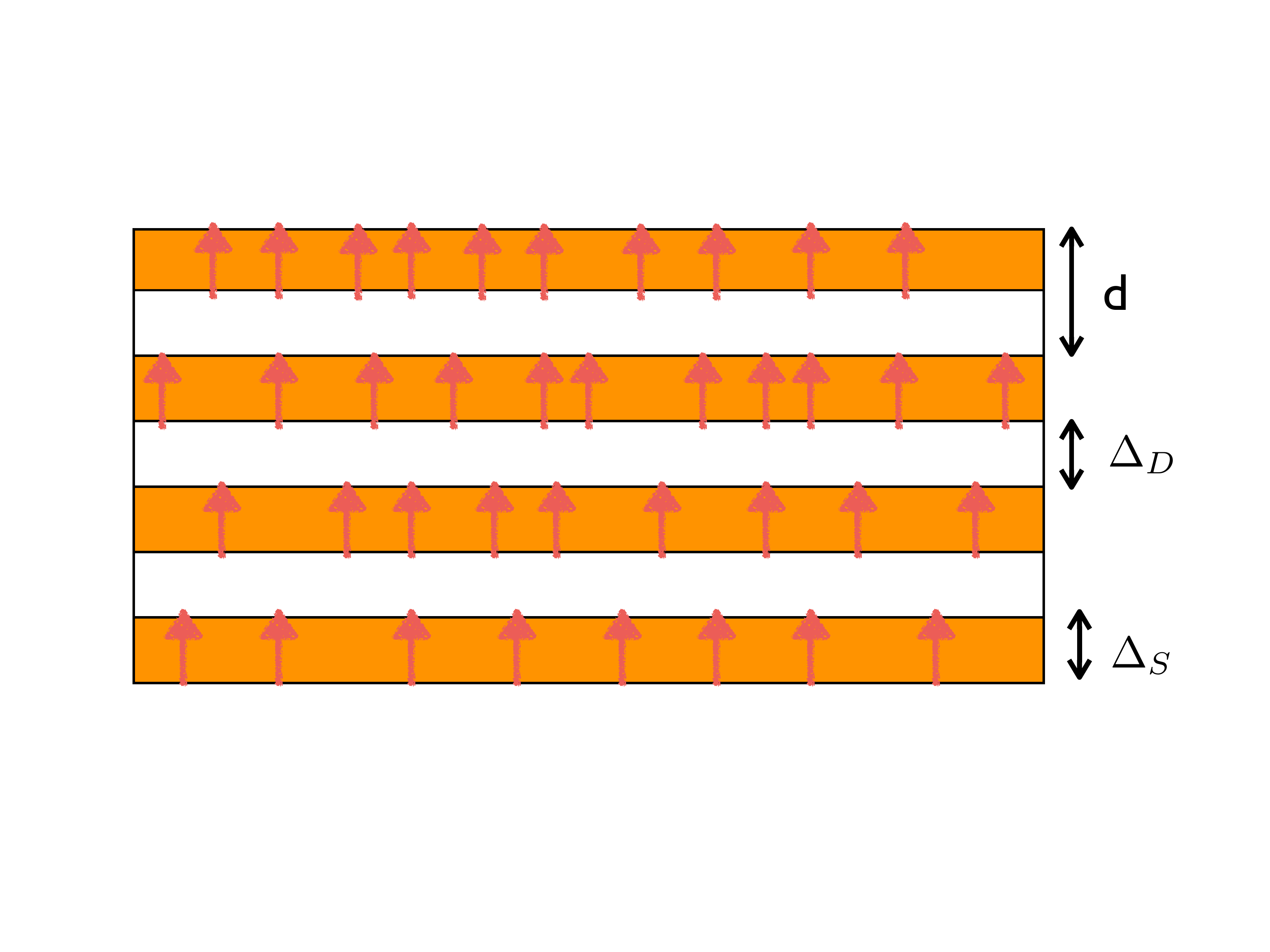}
\vspace{-2cm} 
\caption{Coupled-layer construction of an elementary Weyl semimetal. Magnetically-doped TI layers are coupled 
through insulating spacers. The tunneling amplitude between neighboring TI layers is $\Delta_D$.}
\label{fig:2}
\end{figure}
This implies that this system exhibits a quantum Hall plateau transition from $\sigma_{xy} = 0$ to 
$\sigma_{xy} = e^2/h$ as the ratio of $b/\Delta_s$ is varied and taking both $b$ and $\Delta_S$ to be positive for concreteness.
In other words, in 2D there exists a direct transition between a topological insulator with $\sigma_{xy} = e^2/h$ and a normal 
insulator with $\sigma_{xy} = 0$.
The critical point between the two is described by a massless 2D Dirac Hamiltonian. 

\begin{figure}[t]
\vspace{-2cm}
\includegraphics[width=\columnwidth]{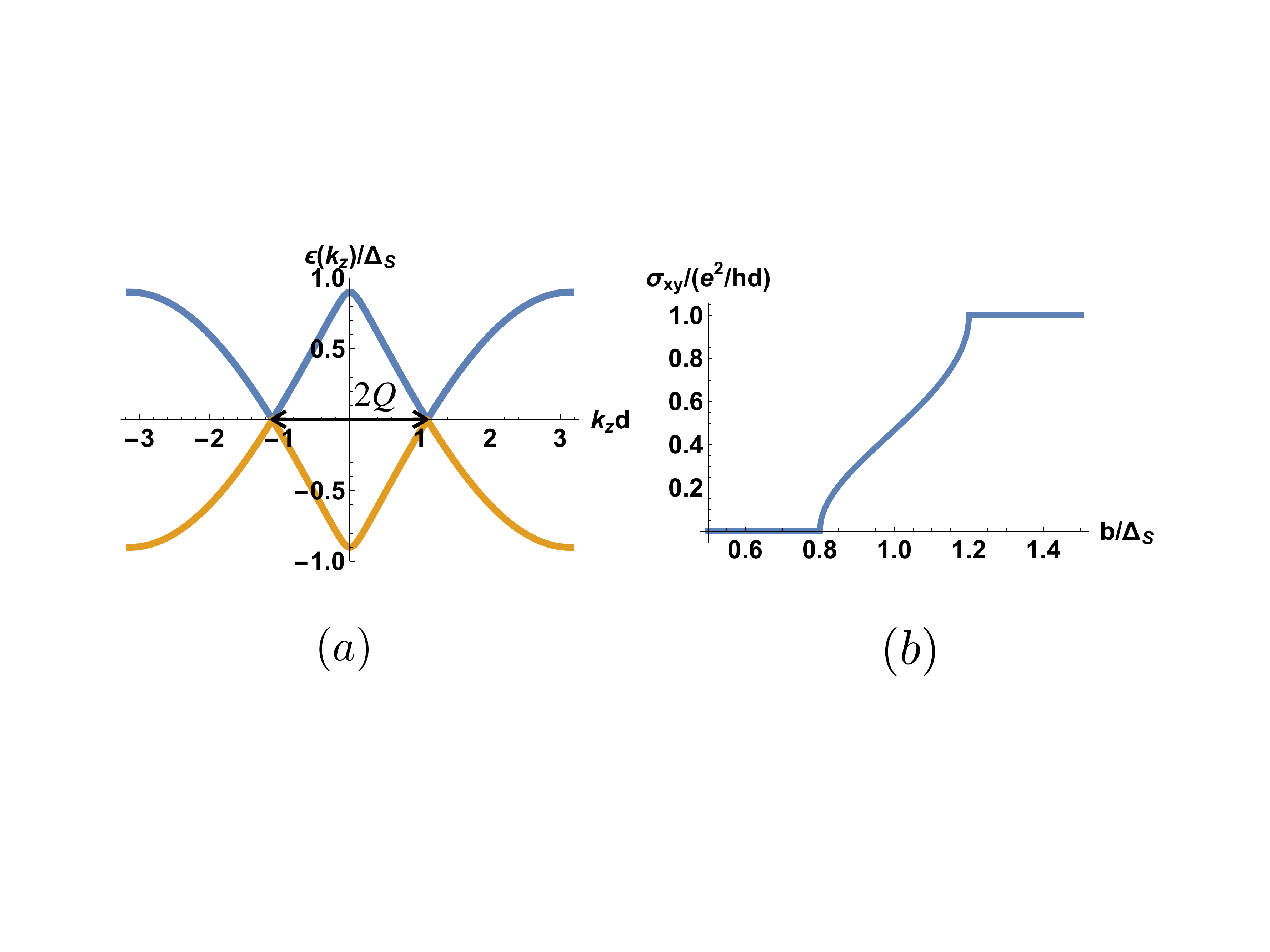}
\vspace{-2cm}
  \caption{(a) Electronic structure of the simplest Weyl semimetal, with two nodes of opposite chirality, separated by 
  a distance of $2 Q$ along the $z$-axis in momentum space. (b) The corresponding anomalous Hall conductivity as 
  a function of $b/\Delta_S$, showing a broadened plateau transition. Weyl semimetal is an intermediate gapless phase 
  between the quantum anomalous Hall and ordinary insulators.}
  \label{fig:3}
\end{figure}

Suppose we now make a stack of such 2D layers, exhibiting QAHE, as shown in {\bf Figure~\ref{fig:2}}. 
Let individual layers be separated by insulating 
spacers, such that the amplitude for tunneling between the adjacent surfaces of neighboring QAHE layers is $\Delta_D$, which 
we will also take to be positive. 
The Hamiltonian, that describes this system, takes the form
\beqa
\label{eq:7}
H&=&\hbar v_F \tau^z (\hat z \times \bsigma) \cdot \bk + [\Delta_S + \Delta_D \cos(k_z d)] \tau^x \nonumber \\
&-&\Delta_D \sin(k_z d) \tau^y + b \sigma^z, 
\eeqa
where $d$ is the period of the resulting superlattice heterostructure in the $z$-direction. 
Making the same unitary transformation as above and partially diagonalizing the resulting Hamiltonian, we obtain
\beq
\label{eq:8}
H_r = \hbar v_F (\hat z \times \bsigma) \cdot \bk + m_r(k_z) \sigma^z, 
\eeq
where $m_r(k_z) = b + r \sqrt{\Delta_S^2 + \Delta_D^2 + 2 \Delta_S \Delta_D \cos(k_z d)} \equiv b + r \Delta(k_z)$. 
Now we see that the quantum Hall plateau transition we discussed before as a function of $b/\Delta_S$, may now happen 
in momentum space as $k_z$ is swept through the BZ. 
Indeed $m_-(k_z)$ will change sign at $k_z^{\pm} = \pi/d \pm Q$, where 
\beq
\label{eq:9}
Q = \frac{1}{d} \arccos\left(\frac{\Delta_S^2 + \Delta_D^2 - b^2}{2 \Delta_S \Delta_D}\right). 
\eeq
At $\bk  = (0, 0, k^{\pm}_z)$, the two nondegenerate bands, corresponding to the eigenvalue $r = -$ touch each other, 
i.e. these are locations of two Weyl nodes, see {\bf Figure~\ref{fig:3}}. 

The nodes exist as long the spin splitting $b$ is in the interval between two critical values 
$b_{c1} < b < b_{c2}$, where $b_{c1} = |\Delta_S - \Delta_D|$ and $b_{c2} = \Delta_S + \Delta_D$. 
When $b < b_{c1}$ the system is an ordinary insulator with $\sigma_{xy} = 0$, while when $b > b_{c2}$ it is a 3D quantum anomalous Hall insulator with $\sigma_{xy} = e^2/h d$.
In between the heterostructure is in the intermediate Weyl semimetal phase with 
\beq
\label{eq:10}
\sigma_{xy} = \frac{2 Q}{2 \pi} \frac{e^2}{h},
\eeq
which depends only on the distance between the Weyl nodes in momentum space and varies continuously between 
$0$ and $e^2/hd$.
Thus, unlike in 2D, in 3D a direct transition between a TI with nonzero quantized Hall conductivity 
and a normal insulator with zero Hall conductivity does not exist. 
The transition instead proceeds through an intermediate gapless Weyl semimetal phase. 
The system, described above, constitutes the simplest potential realization of a topological semimetal. Most of the properties of Weyl and Dirac 
semimetals may be understood by studying this system. 

\section*{Chiral anomaly}
\label{sec:3}
This section describes basics of the topological response of Weyl and other point node semimetals. 
This response may be viewed as a consequence of the chiral anomaly, a phenomenon, originally discovered in the 
particle physics context by Adler, Bell and Jackiw. 
To understand the connection to the chiral anomaly, let us note that the Hall conductivity of a magnetic Weyl semimetal Eq.~\eqref{eq:10}
may be expressed as the following contribution to the imaginary-time action of the electromagnetic field
\beq
\label{eq:11}
S =  i \frac{2 Q}{4 \pi} \frac{e}{\Phi_0} \int d \tau d^3 r \epsilon_{z \nu \alpha \beta} A_{\nu} \partial_{\alpha} A_{\beta}, 
\eeq 
where $A_{\mu}$ is the gauge potential of the electromagnetic field, $\Phi_0 = h/e$ is the magnetic flux quantum, and the Hall conductivity is obtained by varying $S$ with respect to $A_x$
\beq
\label{eq:12}
j_x = - \frac{\delta S}{\delta A_x} = \frac{2 Q}{2 \pi} \frac{e^2}{h} E_y. 
\eeq

To make a connection with the chiral anomaly, we need to first remind ourselves of some basic notions of the crystal elasticity theory. 
A perfect 3D crystal may be described in terms of three families of crystal planes, defined by the relation
\beq
\label{eq:13}
\theta^i(\bR, t) = \bb^i \cdot \bR = 2 \pi n^i, 
\eeq
where $i = 1, 2, 3$, $\bR$ are the Bravais lattice vectors, $\bb^i$ are the three primitive reciprocal lattice vectors and $n^i$ are integers. 
In a distorted crystal, $\theta^i(\tilde \bR, t) = 2 \pi n^i$ continues to hold, where $\tilde \bR$ are the actual lattice site positions, but 
$\partial_j \theta^i = b^i_j$ no longer holds. Instead $\partial_j \theta^i$ may be viewed as components of local coordinate basis vectors of the reciprocal 
space in a distorted crystal. 
It is then useful to view 
\beq
\label{eq:14}
e^i_{\mu} = \frac{1}{2 \pi} \partial_{\mu} \theta^i, 
\eeq
which may have temporal in addition to spatial components, as crystal translational symmetry ``gauge fields". 
In a crystal without dislocations, $d e^i = \frac{1}{2}(\partial_{\mu} e^i_{\nu} - \partial_{\nu} e^i_{\mu}) dx^{\mu} dx^{\nu} = 0$, as follows from Eq.~\eqref{eq:14}. 
If a dislocation with a Burgers vector along $\bb^i$ is present, the integral around a loop, enclosing the dislocation line, is $\oint e^i = 1$. 
Everywhere outside of the dislocation line, however, we may still take $d e^i = 0$, which will be assumed henceforth. 

The point of this exercise in the crystal elasticity theory, is that Eq.~\eqref{eq:11} may now be rewritten as
\beqa
\label{eq:18}
S&=&i \frac{1}{2} \frac{e}{\Phi_0} \frac{2 Q}{2\pi/d} \int d \tau d^3 r \epsilon_{\mu \nu \alpha \beta} e^z_{\mu} A_{\nu} \partial_{\alpha} A_{\beta} \nonumber \\
&=&i \frac{\lambda}{2} \frac{e}{\Phi_0} \int e^z \wedge A \wedge dA. 
\eeqa
Here $\lambda = 2 Q/ (2\pi/d)$ is the dimensionless magnitude of the separation between the Weyl nodes in units of the reciprocal lattice vector. 
This looks like a standard topological term, since it involves only gauge fields and a universal coefficient, except for the fact that the coefficient is
not quantized, since $\lambda$ is not an integer and is continuously tunable. 
This is exactly analogous to the noninteger value of the Luttinger filling parameter $\nu$ in metals, which requires Fermi surface of gapless modes to be 
present. 
In a Weyl semimetal $\nu$ is integer, but a noninteger value of the parameter $\lambda$ still requires gapless modes to be present, but in the 
form of point band-touching nodes rather that a Fermi surface. 

To connect Eq.~\eqref{eq:18} with the standard chiral anomaly we now note the following. 
If we focus only on the low-energy modes, Weyl semimetal appears to have a larger symmetry group than it actually has. 
Its true microscopic symmetry group is $U(1) \times \mathbb{Z}$, where $U(1)$ corresponds to the electric charge conservation and 
$\mathbb{Z}$ to lattice translations (other discrete crystal symmetries, such as rotations, are irrelevant here). 
At low energies, however, the symmetry group appears to be $U(1)_L \times U(1)_R$, which corresponds to separate conservation of Weyl fermions
of left (L) and right (R)-handed chirality. 
This symmetry group, if it were a true microscopic symmetry, would be anomalous, in the sense that it could not be realized in any 3D lattice model, but 
could only appear on the surface
of a four-dimensional (4D) topological insulator, described by the following topological field theory
\beq
\label{eq:19}
S = \frac{i}{6} \frac{e}{\Phi_0^2} \int (A_R \wedge d A_R \wedge d A_R - A_L \wedge d A_L \wedge d A_L),
\eeq
where $A_{R,L}$ are $U(1)$ gauge fields, corresponding to the separate $U(1)_{R, L}$ symmetries. 
We now write $A_R = A + \tilde Q \tilde A$ and $A_L = A - \tilde Q \tilde A$, where $\tilde A$ is a chiral $U(1)$ gauge field, which 
couples antisymmetrically to fermions of opposite chirality and $\tilde Q$ is the corresponding ``charge". 
This gives
\beq
\label{eq:20}
S = i \tilde Q \frac{e}{\Phi_0^2} \int \tilde A \wedge d A \wedge d A.
\eeq
On the 3D boundary of this 4D topological insulator, Eq.~\eqref{eq:20} becomes
\beq
\label{eq:21}
S = i \tilde Q \frac{e}{\Phi_0^2} \int \tilde A \wedge A \wedge d A.
\eeq
If we now identify $\tilde Q = Q d = \pi \lambda$, i.e. the dimensionless magnitude of the momentum of the two Weyl points 
(i.e. translational symmetry charge) and 
$\tilde A = e^z \Phi_0/ 2\pi$, we get precisely Eq.~\eqref{eq:18}, which describes topological response of the physical 3D Weyl semimetal. 
The reason Eq.~\eqref{eq:18} can exist in a stand-alone 3D system, rather than a boundary of a 4D topological insulator, is that the physical translational 
symmetry group is $\mathbb{Z}$ rather than $U(1)$ and, moreover, translation is not a truly ``internal" symmetry, like chiral charge conservation. 
Nevertheless, formally Eq.~\eqref{eq:18} and \eqref{eq:21} are identical, which does lead to a number of unique observable phenomena in topological semimetals, which are usually described as consequences of the chiral anomaly. These are discussed in the following section. 
\section*{Topological magnetotransport phenomena in semimetals}
\label{sec:4}
This section describes some of the observable phenomena, which are encoded in Eq.~\eqref{eq:18}. 
If we vary the action with respect to $A_x$, $A_y$ and $e^z_z$, this gives the Hall conductance per atomic plane
\beq
\label{eq:22}
G_{xy} = \lambda \frac{e^2}{h}, 
\eeq
which in a perfect crystal is equivalent to the Hall conductivity, given by Eq.~\eqref{eq:10}. 
Expressing this in terms of conductance per atomic plane eliminates the dependence on a nonuniversal 
lattice constant $d$, leaving only the pure number $\lambda$. 
A noninteger value of the Hall conductance per atomic plane in units of $e^2/h$ is impossible in a gapped insulator (unless it has topological order), 
which is another way to see that gapless modes must be present in a Weyl semimetal. 

The noninteger Hall conductance of Eq.~\eqref{eq:22} is a property that is specific to magnetic Weyl semimetals with broken TR symmetry. 
A more universal property, common, in some form, to all point-node topological semimetals, is revealed by looking at Eq.~\eqref{eq:18} from a different angle. 
Suppose a straight magnetic flux tube along the $z$-direction, carrying a quantum of flux $\Phi_0$, is inserted into the sample. 
Eq.~\eqref{eq:18} tell us that the action, describing this one-dimensional (1D) magnetic flux tube, is given by
\beq
\label{eq:23}
S =  i e  \lambda \int e^z \wedge A. 
\eeq
This action describes a 1D metal, which has a noninteger charge per unit cell of 
\beq
\label{eq:24}
\rho d = i \frac{\delta S}{\delta A_0 \delta e^z_z} = e \lambda,
\eeq
where $\rho$ is the charge density. 
The Fermi momentum of this metal is 
\beq
\label{eq:25}
k_F = \pi \rho  = \frac{\pi \lambda}{d} = Q, 
\eeq
which means that the 1D band dispersion crosses zero energy at the locations of the band-touching nodes of the 3D Weyl semimetal. 
\begin{figure}[t]
\includegraphics[width=\columnwidth]{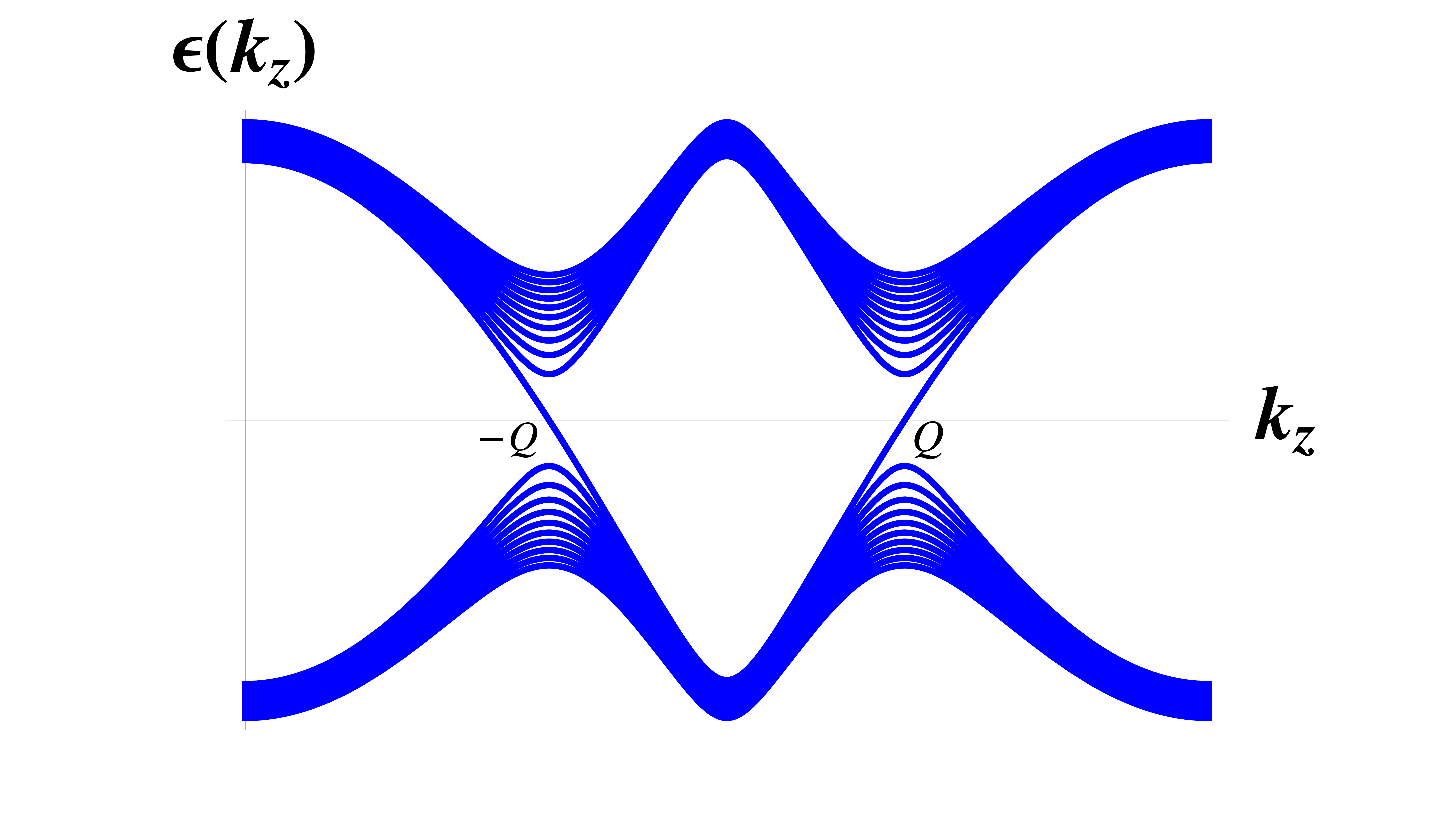}
  \caption{Landau levels of a Weyl semimetal with a pair of Weyl nodes at $k_z = \pm Q$.}
  \label{fig:4}
\end{figure}

This is a simple demonstration of a fundamental property, shared (with some modifications) by all 3D point-node topological semimetals, 
namely the existence of a special Landau level, or more generally levels, which cross the Fermi energy at the locations of the band-touching nodes, 
see {\bf Figure~\ref{fig:4}}. 
In a magnetic Weyl semimetal, discussed above, a magnetic field $B$, applied along the $z$-direction, generates 
\beq
\label{eq:26}
N_{\Phi} = \frac{B L^2}{\Phi_0},
\eeq
1D channels, i.e. one per every flux quantum, which corresponds to a magnetic-field-dependent charge density of 
\beq
\label{eq:27}
\rho = \frac{e \lambda}{d} \frac{B}{\Phi_0}. 
\eeq
According to the Streda formula, this corresponds to the Hall conductivity 
\beq
\label{eq:28}
\sigma_{xy} = \frac{\partial \rho}{\partial B} = \frac{e^2}{h} \frac{\lambda}{d}, 
\eeq
which agrees with Eq.~\eqref{eq:10}. 

Aside from the Hall conductivity of magnetic Weyl semimetals, this Landau level is responsible for all the special magnetotransport 
phenomena in 3D point-node topological semimetals. 
Consider diffusive transport equations for a Weyl semimetal with a single pair of opposite-chirality nodes, for simplicity. 
There are two conserved quantities: total electron density $n = n_R + n_L$ and chiral electron density $n_c = n_R - n_L$, where 
$n_{R, L}$ are the densities of right ($R$) and left ($L$)-handed electrons. 
The transport equations for $n$ and $n_c$ are given by
\beqa
\label{eq:29}
\frac{\partial n}{\partial t}&=&D \bnabla^2 (n + g V) + \bGamma \cdot \bnabla n_c, \nonumber \\
\frac{\partial n_c}{\partial t}&=&D \bnabla^2 n_c - \frac{n_c}{\tau_c} + \bGamma \cdot \bnabla (n + gV). 
\eeqa
Here $V$ is the external electrostatic potential, $D \sim v_F^2 \tau$ is the diffusion constant, $\tau$ is the momentum relaxation time, which depends on the density of states $g$ and the strength of the impurity scattering. $\tau_c \gg \tau$ is the chirality relaxation time, which is assumed to be much longer than the momentum relaxation time. 
The density of states $g$ is taken to be finite even in the absence of the external magnetic field $\bB$. This may be due to the Fermi energy 
not exactly coinciding with the location of the Weyl nodes, which will generically be the case in the presence of impurities. 
In this case $g = \epsilon_F^2/ (\pi^2 \hbar^3 v_F^3) = k_F^2/(\pi^2 \hbar v_F)$ for a pair of Weyl nodes. 
A finite density of states may also be directly induced by strong enough disorder even when $\epsilon_F = 0$. 

$\bGamma$ is a new transport coefficient, related to the chiral anomaly, given by
\beq
\label{eq:30}
\bGamma = \frac{\hat B}{2 \pi^2 \hbar \ell_B^2 g} \sim \frac{v _F}{(k_F \ell_B)^2} \hat B, 
\eeq
where $\ell_B = \sqrt{\hbar/ e B}$ is the magnetic length and $\hat B$ is a unit vector in the direction of the magnetic field. 
Since the total electron number is exactly conserved, we may read off the expression for the electric current from the first of Eq.~\eqref{eq:29}
\beq
\label{eq:31}
\bj = \frac{\sigma}{e} \bnabla \mu + e g \mu_c \bGamma, 
\eeq
where $\sigma = e^2 g D$ is the diagonal conductivity and $\mu, \mu_c$ are the electrochemical potentials, associated with the total and chiral 
particle densities. 
The first term in Eq.~\eqref{eq:31} is the standard Ohm's law. The second term describes an extra contribution to the electrical conductivity, which 
 arises from the chiral anomaly, namely the $N_{\Phi}$ 1D conduction channels, which are opened up by the applied field. 
 It is clear that this additional contribution to the current can have a dramatic effect in transport, since it is associated with first derivative terms 
 in the transport equations Eq.~\eqref{eq:29}, while the standard drift-diffusion terms which involve second derivatives. 
 
 To quantify this effect, it is useful to find the eigenmodes of the transport equations, which are given by
\beq
\label{eq:32}
\omega_{\pm} = \pm \omega_0 - i (D q^2 + 1/2 \tau_c), 
\eeq
where 
\beq
\label{eq:33}
\omega_0 = \sqrt{(\Gamma q)^2 - 1/4 \tau_c^2}. 
\eeq
We now note that the frequency $\omega_0$ is purely imaginary at the smallest momenta when 
\beq
\label{eq:34}
q < \frac{1}{2 \Gamma \tau_c} = \frac{L_a}{2 L_c^2} \equiv \frac{1}{L_*}, 
\eeq
where we have introduced two length scales 
\beq
\label{eq:35}
L_c = \sqrt{D \tau_c},
\eeq
which has the meaning of the chiral charge diffusion length, and 
\beq
\label{eq:36} 
L_a = \frac{D}{\Gamma} \sim \ell (k_F \ell_B)^2, 
\eeq
where $\ell = v_F \tau$ is the mean free path. 
$L_a$ is a magnetic-field-related length scale, distinct from the magnetic length, which arises from the chiral anomaly. 
It is a long hydrodynamic length scale in the weak magnetic field regime, in the sense that $L_a \gg \ell$, but it may still be 
much smaller that either the chiral charge diffusion length $L_c$ or the sample size $L$. 
In fact, the ratio $L_c/L_a$ quantifies the strength of the chiral-anomaly-related transport phenomena, as will be seen below. 

Thus when $q < 1/L_*$ the eigenfrequencies of the transport equations are purely imaginary, which corresponds to ordinary diffusion (nonpropagating) 
modes. 
However, when $q > 1/L_*$ (which may be a very small momentum when the ratio $L_c/L_a$ is large), $\omega_0$ is real, which signals the emergence of a pair of propagating modes in this regime. 
The modes are only weakly damped as long as 
\beq
\label{eq:37}
\omega_0 \approx \Gamma q > D q^2, 
\eeq
which defines the upper limit on the wavevector $q = 1/L_a$, above which the propagating modes disappear. 
The propagating modes thus exist in the interval 
\beq
\label{eq:38}
1/L_* < q < 1/L_a. 
\eeq
This interval is significant when $L_c/ L_a \gg 1$. 
On the other hand, when $L_a > L_c$, propagating modes do not exist for any $q$ and one obtains a pair of standard diffusion modes
\beq
\label{eq:39}
\omega_+ = - i D q^2, \,\, \omega_- = - i D q^2 - i/\tau_c, 
\eeq
which corresponds to independent diffusion of the total and the chiral charge densities. 

The propagating mode of Eq.~\eqref{eq:37} signals the emergence of quasiballistic transport at long enough length scales. 
This may be seen explicitly by solving Eq.~\eqref{eq:29} in the steady state, assuming a uniform sample of linear size $L$, attached to metallic leads 
in the $z$-direction (i.e. the current flows along the magnetic field). One obtains the following expression for the scale-dependent sample conductance
\beq
\label{eq:40}
G(L) = \frac{e^2 N_{\phi}}{h} F(L/L_a, L/L_c), 
\eeq
where the scaling function $F(x,y)$ is given by
\beq
\label{eq:41}
F(x,y) = \frac{(1 + y^2/x^2)^{3/2}}{\frac{y^2}{2 x} \sqrt{1 + y^2/x^2} + \tanh \left(\frac{x}{2} \sqrt{1 + y^2/x^2}\right)}. 
\eeq
This scaling function exhibits crossover behaviors which exactly match the corresponding crossovers in the wavevector dependence 
of the diffusion modes, found above. 

Indeed, when $x \ll y$, which means $L_a \gg L_c$, we have $F(x, y) \approx 2/x$, which gives 
\beq
\label{eq:42}
G(L) \approx e^2 g D L = \sigma L, 
\eeq
which is simply the standard Ohmic conductance, with a small magnetic-field dependent correction, which goes as $(L_c/L_a)^2$, and which we have ignored here for the sake of brevity.
This arises in the regime, in which we have two independent diffusion modes, given by Eq.~\eqref{eq:39}, corresponding to independent diffusion of the total and the chiral charges. 

On the other hand, when $L_a \ll L_c$, or $x \gg y$, we obtain 
\beq
\label{eq:43}
F(x, y) \approx \frac{1}{y^2/ 2 x + \tanh(x/2)}.
\eeq
This exhibits a regime of quasiballistic conductance with 
\beq
\label{eq:44}
G(L) \approx \frac{e^2 N_{\phi}}{h}, 
\eeq
which is realized when 
\beq
\label{eq:45}
L_a < L < L_*.
\eeq
This corresponds precisely to the range of the wavevectors $q$ in Eq.~\eqref{eq:38}, for which propagating modes exist when $L_a \ll L_c$. 
Thus, one of the observable manifestations of the existence of quasi-1D propagating modes in a topological semimetal is the quasiballistic 
conductance, given by Eq.~\eqref{eq:44}.
Upon further increase of the sample size, a crossover back into the diffusive regime happens when $y^2 > x$, or 
equivalently when the sample size exceeds the length $L_*$
\beq
\label{eq:46} 
L > L_* \gg L_c. 
\eeq
In this regime the conductance is dominated by the field-dependent chiral anomaly contribution, but with $\propto B^2$, instead 
of $\propto B$, dependence
\beq
\label{eq:47}
G(L) = \sigma (L_c/L_a)^2 L. 
\eeq
When $L_a > L_c$ this result still holds in the infinite sample limit, but with Eq.~\eqref{eq:47} only representing 
a subdominant correction to the Ohmic conductance. 
\begin{figure}[t]
\includegraphics[width=\columnwidth]{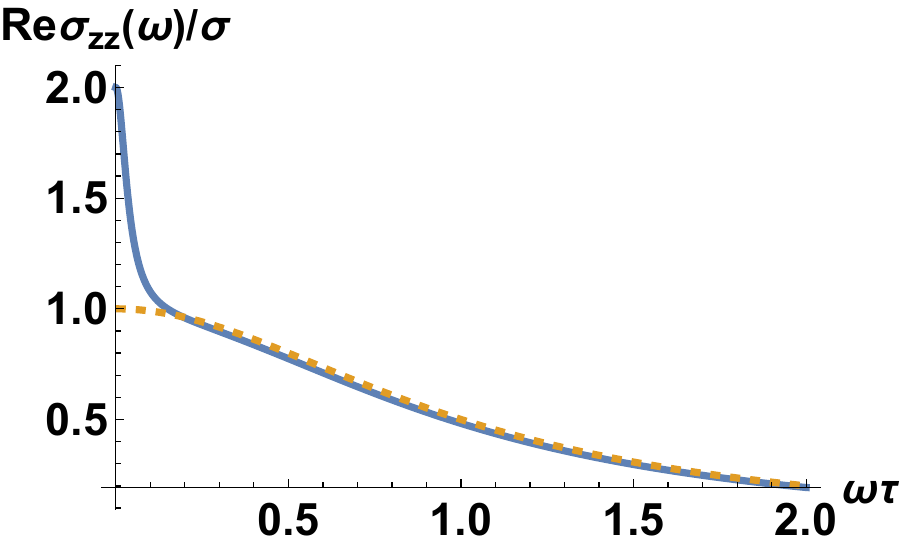}
\caption{Frequency-dependent conductivity for $L_c/L_a = 1$ (solid line) and $L_c/L_a = 0$ (dashed line), and $\tau/\tau_c = 0.04$.}
\label{fig:optical}
\end{figure}

Perhaps the clearest evidence for the chiral anomaly can come from the measurement of the optical magnetoconductivity. 
This may be evaluated using the full frequency and wavevector-dependent density response function $\chi(q, \omega)$, with ladder impurity vertex 
corrections
\beq
\label{eq:48}
\sigma_{zz}(\omega) = - e^2 \lim_{q \rightarrow 0} \frac{i \omega}{q^2} \chi(q, \omega). 
\eeq
A straightforward calculation then gives
\beq
\label{eq:49}
\sigma_{zz}(\omega) = \frac{\sigma}{1 - i \omega \tau} \frac{1 - i \omega \tau_c + (L_c/ L_a)^2}{1 - i \omega \tau_c}.
\eeq
Evaluating the real part, one obtains
\beq
\label{eq:50}
\textrm{Re} \,\sigma_{zz}(\omega) = \frac{\sigma}{1 + \omega^2 \tau^2} \left[1 + \left(\frac{L_c}{L_a}\right)^2 \frac{1 - \omega^2 \tau \tau_c}{1 + \omega^2 \tau_c^2} \right]. 
\eeq
The prefactor in Eq.~\eqref{eq:50} is the standard Drude expression for the optical conductivity of a metal. 
The part in the square brackets is a correction that arises in a topological semimetal as a consequence of the chiral anomaly. 
This correction represents transfer of the spectral weight from high frequencies into a new low-frequency peak, whose width scales with the 
chiral charge relaxation rate $1/\tau_c$, while height is proportional to the ratio $(L_c/L_a)^2$. 
Importantly, Eq.~\eqref{eq:50} satisfies the exact $f$-sum rule
\beq
\label{eq:51}
\int_0^{\infty} d \omega\,\, \textrm{Re}\, \sigma_{zz}(\omega) = \frac{\pi \sigma}{2 \tau}, 
\eeq
which means that the appearance of the new low-frequency peak indeed represents spectral weight transfer, as it should, see {\bf Figure~\ref{fig:optical}}. 
The extra zero-frequency peak in the optical conductivity is a direct evidence of the opening up of a new transport channel (the zeroth Landau level) 
by the applied magnetic field.
\section*{Conclusions}
In conclusion, we have described topological properties of Dirac and Weyl semimetals. These may be viewed as a consequence of the chiral anomaly and 
manifest in observable phenomena: intrinsic anomalous Hall effect with a non-integer Hall conductance per atomic plane; negative longitudinal magnetoresistance and scale-dependent quasiballistic transport; extra magnetic-field-dependent low-frequency peak in the optical conductivity, 
whose width is determined by the chirality relaxation rate. 
\section*{Keywords}
Topological phases of matter, topological semimetals, Dirac and Weyl semimetals, quantum Hall effect. 
\section*{Further Reading}
\begin{enumerate}
\item[] N.P. Armitage, E.J. Mele, and A. Vishwanath, Weyl and Dirac Semimetals in three-dimensional solids, {\em Rev. Mod. Phys. {\bf 90}, 015001 (2018)}. 
\item[] A.A. Burkov and L. Balents, Weyl semimetal in a topological insulator multilayer, {\em Phys. Rev. Lett. {\bf 107}, 127205 (2011)}. 
\item[] D.T. Son and B.Z. Spivak, Chiral anomaly and classical negative magnetoresistance of Weyl metals, {\em Phys. Rev. B {\bf 88}, 104412 (2013)}. \\
\item[]  A.Altland and D. Bagrets, Theory of the strongly disordered Weyl semimetals, {\em Phys. Rev. B {\bf 93}, 075113 (2016)}. 
\item[] A.A. Burkov, Dynamical density response and optical conductivity in topological metals, {\em Phys. Rev. B {\bf 98}, 165123 (2018)}. 
\item[] L. Gioia, C. Wang, and A.A. Burkov, Unquantized anomalies in topological semimetals, {\em Phys. Rev. Research {\bf 3}, 043067 (2021)}. 
\item[] J. Nissinen and G.E. Volovik, Elasticity tetrads, mixed axial-gravitational anomaly and $(3+1)-d$ quantum Hall effect, {\em Phys. Rev. Research 
{\bf 1}, 023007 (2019)}. 
\item[] J. Xiong, S. K. Kushwaha, T. Liang, J. W. Krizan, M. Hirschberger, W. Wang, R. J. Cava, and N. P. Ong, Evidence for the chiral anomaly in the Dirac semimetal Na$_3$Bi, Science {\bf 350}, 413 (2015).
\item[] Q. Li, D. E. Kharzeev, C. Zhang, Y. Huang, I. Pletikosic, A. V. Fedorov, R. D. Zhong, J. A. Schneeloch, G. D. Gu, and T. Valla, Chiral magnetic effect in ZrTe$_5$, Nat Phys {\bf 12}, 550 (2016).
\item[] B. Cheng, T. Schumann, S. Stemmer, and N.P. Armitage, Probing charge pumping and relaxation of the chiral anomaly in a Dirac semimetal, 
Sci. Adv. {\bf 7}, eabg0914 (2021). 
\end{enumerate}
\end{document}